\documentclass[prb,twocolumn,showpacs,amsmath,amssymb]{revtex4}
\usepackage{graphicx}% Include figure files
\usepackage{dcolumn}% Align table columns on decimal point
\usepackage{bm}% bold math
\usepackage[tight]{subfigure}
\usepackage{amsmath}
\usepackage{verbatim}
\usepackage{color}
\begin{document}

\title{Finite-bias conductance anomalies at a singlet-triplet crossing}
\author{Chiara Stevanato, Martin Leijnse, Karsten Flensberg and Jens Paaske}
\affiliation{Center for Quantum Devices, The Niels Bohr Institute, University of Copenhagen, 2100~Copenhagen \O, Denmark}

\begin{abstract}
Quantum dots and single-molecule transistors may exhibit level crossings induced by tuning external parameters such as magnetic field or gate voltage. For Coulomb blockaded devices, this shows up as an inelastic cotunneling threshold in the differential conductance, which can be tuned to zero at the crossing. Here we show that, in addition, level crossings can give rise to a nearly vertical {\it step-edge}, {\it ridge} or even a Fano-like {\it ridge-valley} feature in the differential conductance inside the relevant Coulomb diamond. We study a gate-tunable quasidegeneracy between singlet and triplet ground states, and demonstrate how these different shapes may result from a competition between nonequilibrium occupations and weak (spin-orbit) mixing of the states. Our results are shown to be in qualitative agreement with recent transport measurements on a Mn complex [E. A. Osorio {\it et al.}, Nano Lett {\bf 10}, 105 (2010)]. The effect remains entirely general and should be observable in a wide range of Coulomb blockaded devices.
\end{abstract}
\maketitle

\section{Introduction.}
Low-temperature transport measurements have revealed level crossings in Coulomb blockaded quantum dots and single-molecule transistors, induced either by tuning  gate voltage\cite{Kogan03, Hauptmann08, Roch08a, Osorio10} or external magnetic field\cite{Nygard00, Fasth07, Katsaros10, Jespersen11}. These show up in a bias spectroscopy as inelastic cotunneling thresholds crossing zero, when tuning the control parameter towards the level crossing. If the two different ground states have the same elastic cotunneling amplitudes, the linear conductance shows hardly any trace of the crossing, except perhaps for a peak right at the crossing if the degeneracy gives rise to Kondo-effect~\cite{Hauptmann08, Jespersen11}. In the case of a transition from a singlet to a triplet ground state, however, the elastic cotunneling amplitude can easily be different on the two sides, and the triplet state can even give rise to Kondo effect~\cite{Pustilnik01prb, Kogan03, Roch08a, Osorio10}. The difference in elastic cotunneling leads to a sudden change in the linear conductance at the crossing, which may be smeared by a small mixing of the two competing ground states. In the case of a singlet-triplet crossing, a mixing due to spin-orbit coupling leads to an avoided crossing, which can be observed as a lower bound on the inelastic cotunneling threshold~\cite{Fasth07, Katsaros10, Jespersen11}.

Two recent experiments on widely different single-molecule transistors have shown gate-induced singlet-triplet crossings. These experiments revealed an extra cotunneling line in the differential conductance pinned to the crossing at zero bias and extending up to bias voltages much larger than the other excited spin states, thus demarcating the boundary between two different regions~\cite{Roch08a, Osorio10}. In (the supplementary material to) Ref.~\onlinecite{Osorio10}, see Figs. 6S and 7S, reproduced here in Fig.~\ref{fig:FigExpCuts}, it was pointed out that such a feature could easily arise if the device has sufficiently asymmetric coupling to source and drain. In this  case, even at finite bias voltage the molecule is largely in equilibrium with the strongest coupled electrode. Like at zero bias, a difference in the elastic cotunneling amplitude may therefore show up as a step edge extending into the finite bias region.
\begin{figure}[t]
\includegraphics[width=0.35\textwidth]{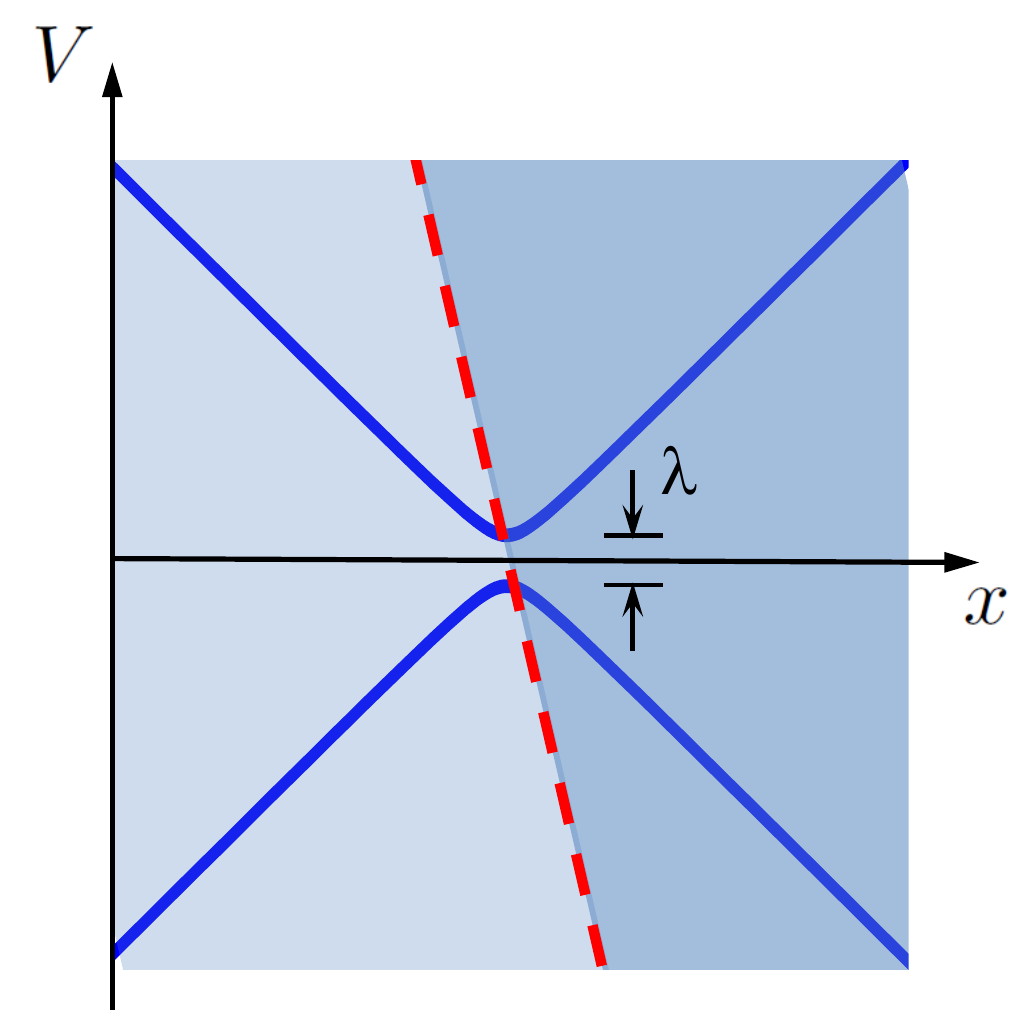}
\caption{\label{fig:LevCross}Sketch of differential cotunneling conductance as a function of bias voltage, $V$, and control parameter, $x$. Thick blue (solid) lines indicate inelastic cotunneling thresholds with an avoided crossing near zero bias, caused by a finite mixing, $\lambda$, of the two involved states. The red (dashed) line indicates the line along which the levels cross (possibly resulting in a conductance peak), including the possibility of a slight bias dependence of the level splitting. The two different sides of this line may have a difference in $dI/dV$ (indicated by different shades of blue) due to a difference in elastic cotunneling amplitudes for the two states.}
\end{figure}
This scenario is sketched in Fig.~\ref{fig:LevCross}, with differential conductance as a function of bias voltage $V$ and control parameter $x$, which in these two experiments is gate voltage. In both experiments this line has a finite slope, implying that the bias voltage somehow influences the tuning of the crossing as part of the total effective gating of the relevant states. In this case, the step edge is accompanied by a peak deriving from differentiating a slanted step in current. In Fig.~\ref{fig:LevCross} the difference in conductance is reflected by two different shades of blue, and a conductance peak may be observed along the border marked by a red (dashed) line. This behavior can indeed be seen from Fig. S6a in Ref.~\onlinecite{Roch08a}, but the corresponding feature observed in Ref.~\onlinecite{Osorio10} revealed a surprising dip-peak Fano-like structure, which remained unexplained.

In this paper, we show that the dip-peak lineshape observed in Ref.~\onlinecite{Osorio10} can arise from a competition between nonequilibrium effects and spin-orbit mixing. In general, a mixing, $\lambda$, of the two states can lead to an avoided crossing as indicated in Fig.~\ref{fig:LevCross}. However, in the experiment \cite{Osorio10} the spin-orbit coupling is weak enough that no avoided crossing is observed in the inelastic cotunneling, i.e., it is smaller than the tunnel-induced broadening of the cotunneling lines. Since the line is only weakly bias dependent (nearly vertical), its detailed line shape still provides otherwise inaccessible information about the spin-orbit coupling. We shall restrict our attention to the singlet-triplet crossing observed in Ref.~\onlinecite{Osorio10}, but the basic mechanism is of general validity and this phenomenon could be observed in a variety of cotunnel junctions.

The paper is organized as follows. In Sect.~\ref{sec:model} we introduce the double-dot model which will be considered throughout the paper. In Sect.~\ref{sec:rateeq} we present the results of a perturbative generalized master equation calculation of transport as a function of gate and bias voltage. This calculation reveals the singlet-triplet boundary in finite-bias cotunneling, and shows that it should connect to kinks in the Coulomb diamond edges arising from sequential tunneling. In Sect.~\ref{sec:kondo}, we focus on the interior of the Coulomb diamond and recalculate the inelastic cotunneling spectrum within an effective exchange cotunneling model. The effective cotunneling model is first established from a Schrieffer-Wolff transformation of the original double-dot model, and the features of Sect.~\ref{sec:rateeq} are reproduced. Finally, we include a singlet-triplet mixing in the form of a spin-orbit induced Dzyaloshinskii-Moriya term, and demonstrate how this causes a nonequilibrium population inversion which in turn gives rise to the peak-dip structure observed experimentally in Ref.~\onlinecite{Osorio10}.

\section{Double dot model}\label{sec:model}

\begin{figure}
\includegraphics[width=0.4\textwidth]{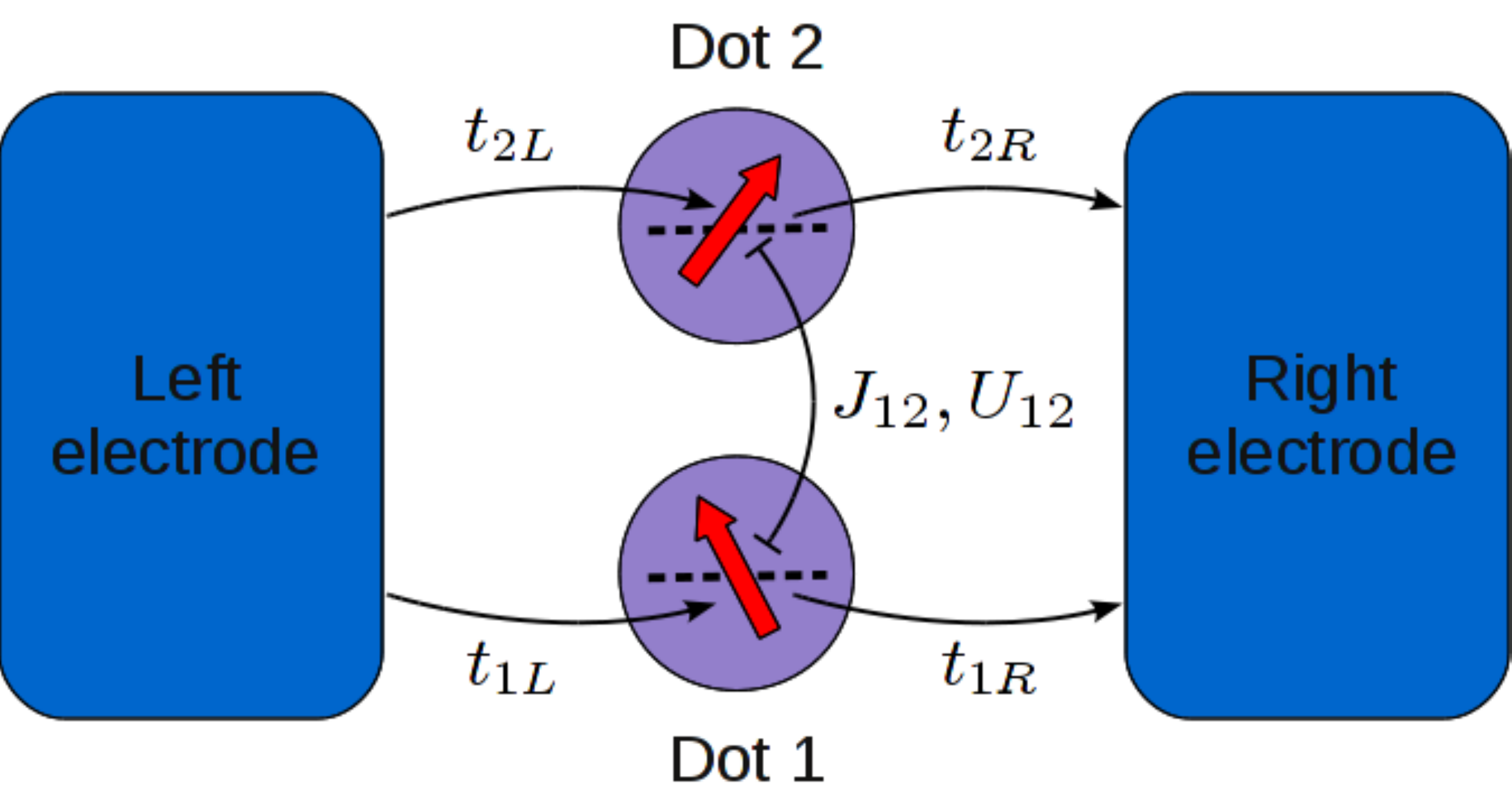}
\caption{\label{fig:DDmodel}(Color online) Sketch of double dot model. Electrons on different dots interact via their charge (Coulomb repulsion, $U_{12}$) as well as via their spin (exchange, $J_{12}$). Tunneling between electrode $r$ and  dot $i$ takes place with amplitude $t_{ir}$. The gate electrode is not shown.}
\end{figure}

As a simple model which can undergo a transition between singlet and triplet groundstates, we consider an interacting double quantum dot, where the two dots are connected in parallel to a left ($L$) and a right ($R$) electrode and capacitively coupled to a gate electrode, see Fig.~\ref{fig:DDmodel}. The Hamiltonian is given by $H = H_{DD} + H_R + H_T$, where
\begin{align}
\label{eq:HD}
	H_{DD} &= \sum_{i \sigma} \epsilon_{i \sigma} n_{i \sigma} + \sum_{i} U_i n_{i \uparrow} n_{i \downarrow} + U_{12} n_1 n_2 \nonumber \\
	    &+ J_{12} \mathbf{S}_1 \cdot \mathbf{S}_2 + B \left( S_1^z + S_2^z \right), \\
\label{eq:HR}
	H_R &= \sum_{r k \sigma} \epsilon_{r k} n_{r k \sigma}, \\
\label{eq:HT}
	H_T &= \sum_{r k i \sigma} t_{r i} c_{r k \sigma} d_{i \sigma}^\dagger + h.c.
\end{align}
The double dot system is described by $H_{DD}$, where $\epsilon_{i \sigma}$ is the energy of the single-particle orbital on dot $i$, which has number operator $n_i = \sum_\sigma n_{i \sigma} = \sum_{\sigma} c_{i \sigma}^\dagger c_{i \sigma}$ for spin projection $\sigma = \uparrow,\downarrow$. $U_i$ and $U_{12}$ are the onsite and inter-dot Coulomb charging energies, respectively. $\mathbf{S}_i$ is the spin on dot $i$ (where $\mathbf{S}_i = \mathbf{0}$ if dot $i$ is empty or doubly occupied). There is an interaction between the spins on the different dots, which can be dominated by either superexchange ($J_{12} > 0$) or Hund's rule coupling ($J_{12} < 0$), and arises as an effective low-energy description of a more complicated multi-orbital double dot model, where orbitals on different dots are partially overlapping~\cite{Osorio10}. $B$ is the magnetic field, which we assume to be applied in the $z$-direction and therefore coupled to the $z$-projection of the dot spins, $S_i^z$.

$H_R$ describes non-interacting electrons in reservoir $r = L,R$, with energy $\epsilon_{r k}$ and number operator $n_{r k \sigma} = c_{r k \sigma}^\dagger c_{r k \sigma}$. Tunneling between the dots and the electrodes is described by $H_T$. For simplicity we assume that the tunnel amplitudes, $t_{r i}$, are independent of $k$ and that the density of states in the electrodes, $\rho_r$, is energy independent. Then also the tunnel rates, $\Gamma_{r i} = 2\pi\rho_r |t_{r i}|^2$, which set the inverse time-scale for resonant single electron tunneling (SET), are energy-independent.

\section{Full gate and bias spectroscopy}\label{sec:rateeq}

We calculate the nonequilibrium reduced density matrix of the double quantum dot, and the current flowing between the electrodes, $I$, using a generalized master equation (GME) approach~\cite{Leijnse08a, Koller10}, based on the real-time transport theory~\cite{Koenig96a}. This approach treats all the local interactions ($U_i, U_{12}, J_{12}$) non-perturbatively. The tunneling between the dots and electrodes is treated perturbatively up to fourth order in $H_T$, which includes all coherent one- and two-electron tunnel processes, such as SET, elastic and inelastic cotunneling~\cite{DeFranceschi01}, pairtunneling~\cite{Koch05c, Leijnse09a}, and tunnel-induced level broadening and shift. The Kondo effect is not included at this level of perturbation theory, which is therefore applicable only at temperatures much larger than the Kondo temperature. In fact, for the theory to be applicable also in the SET regime, the thermal energy scale must dominate over the tunnel broadening, $T \gg \Gamma_{ri}$.
%If, in addition, the single-particle orbitals are non-degenerate
%on the energy scale of the tunnel broadening, $|\epsilon_1 - \epsilon_2| \gg \Gamma_{ri}$, the only degenerate states are protected by selection rules. In this case, the non-diagonal elements of the density matrix, $\rho_{ab}$, can be calculated as described in Refs.~\onlinecite{Leijnse08a, Koller10} within a lowest order expansion in $\Gamma_{r i} / \Delta E_{ab}$, where $\Delta E_{ab}$ is the
%energy difference of the many-body eigenstates $a$ and $b$.
%\hl{What is the $S-T_0$ issue at $B>0$?}

\begin{figure}
\includegraphics[width=0.5\textwidth]{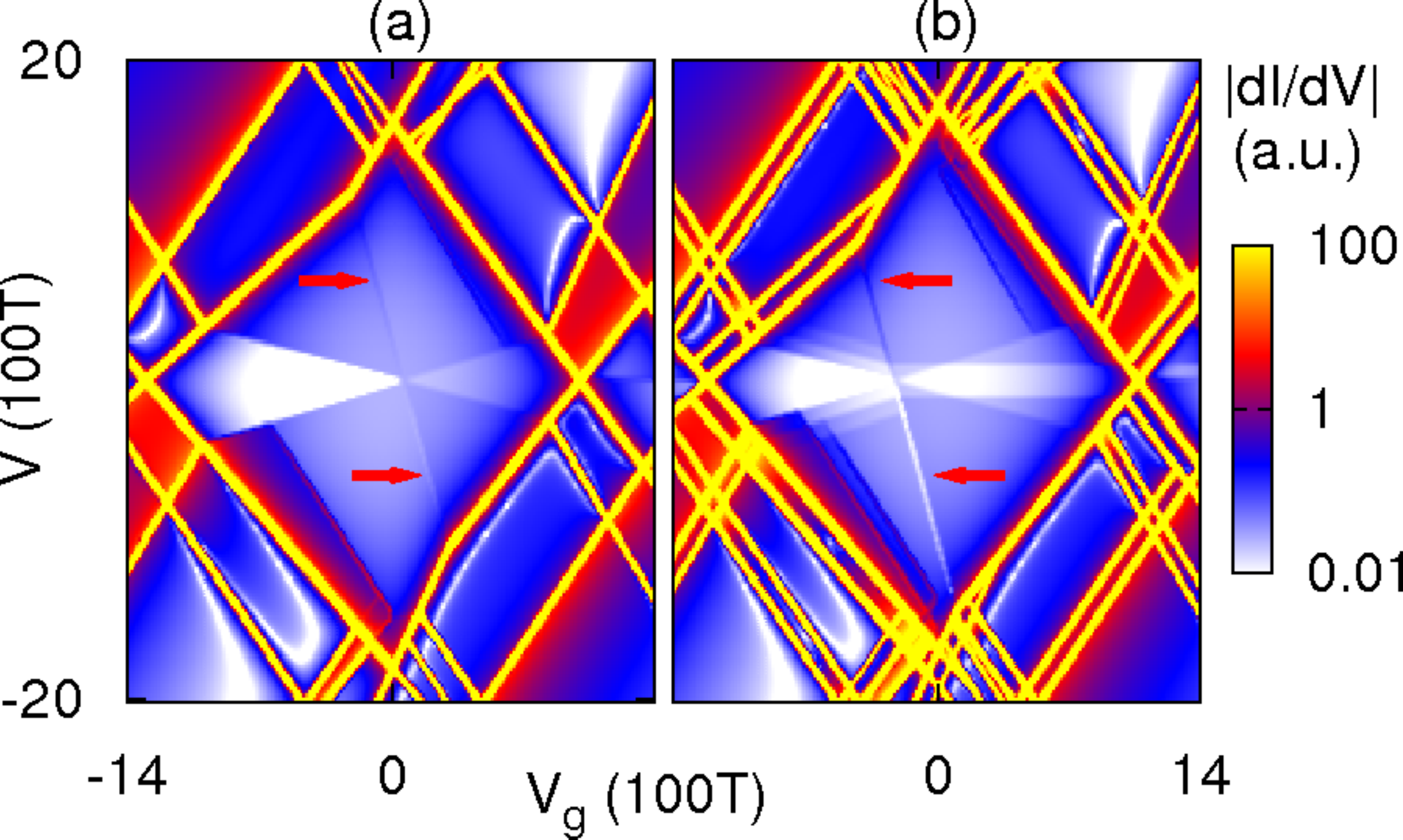}
\caption{\label{fig:diamonds}(Color online) Results of GME calculation presented as conductance maps, $|dI/dV|$ plotted on a logarithmic color scale as a function of $V$ and $V_g$ (the absolute value is plotted since negative values, occuring in certain regions because of the tunnel-rate asymmetry, are not well represented on a logarithmic scale). The red arrows point out the singlet-triplet line. Here the tunnel amplitudes are $t_{1R} = 30, t_{1L} = 0.7, t_{2R} = 0.8, t_{2L} = 0.5$ (arbitrary units), the Coulomb charging energies are $U_{11} = U_{22} = 2000 T$, $U_{12} = 400 T$, and we choose $\epsilon_2 = \epsilon_1 + 2T$. $B = 0$ in (a) and $B = 50 T$ in (b). }
\end{figure}

The results of the GME calculations are shown in Fig.~\ref{fig:diamonds}, represented as two-dimensional conductance maps, with the absolute value of the differential conductance, $|dI/dV|$, plotted on a color scale as a function of bias voltage, $V$, and gate voltage, $V_g$.
Because of capacitive effects, the energy difference between states with $N = \sum_i n_i$ and $N-1$ electrons on the dot is $\propto - N V_g$, where we have set the proportionality constant (gate coupling) to one for simplicity. We assume equal capacitive couplings to left and right electrodes and apply the bias voltage symmetrically with chemical potentials $\mu_{L,R} = \pm V/2$, in which case the bias voltage does not result in direct capacitive energy-shifts. However, as was shown in Ref.~\onlinecite{Osorio10}, $J_{12}$ may depend on both $V$ and $V_g$ (for simplicity we assume a linear dependence).
Therefore the $N=2$ groundstate will change from singlet to triplet along a line in the $(V,V_g)$ plane.
%(note that $\mathbf{S}_1 \cdot \mathbf{S}_2 \neq 0$ only for $N=2$).
%Taking the crossing point to be the zero of our gate voltage axis, we thus assume an exchange coupling of the form:
%\begin{align}
%J_{12}=-\alpha V_g - \beta V,
%\end{align}
%in terms of two dimensionless numbers, $\alpha$ and $\beta$.

Figure~\ref{fig:diamonds}(a) shows the conductance within a voltage range where $N = 2$ in equilibrium.
Already the rather simple model Hamiltonian~(\ref{eq:HD}) has a complicated many-body eigenspectrum and gives rise to a rich conductance map, but we focus here on the features originating from the transition between singlet and triplet ground states. The strong (yellow) conductance lines with a large gate dependence mark the onset of SET. The crossing of such lines at zero bias voltage indicates a charge-degeneracy point. 
The lines emmanating from the zero-bias crossing points at the left (right) indicates the onset of SET involving the $N=1$ ($N=3$) charge states.
Inside the central region (Coulomb diamond), the charge is fixed to $N=2$, and the current is dominated by coherent two-electron cotunneling processes, where an electron is effectively transferred from one electrode to the other, involving only virtual occupation of either the $N=1$ or $N=3$ charge states. A cotunneling process which does not transfer any net energy to or from the double dot is called elastic and gives rise to a constant background conductance. In an inelastic cotunneling process~\cite{DeFranceschi01}, in contrast, energy is transferred to the double dot, which becomes possible when that energy can be supplied by the bias voltage. This gives rise to conductance features whenever the bias voltage equals an excitation energy of the double dot system. In first approximation the conductance feature is a step, but the joint effects of nonequilibrium population of excited states and Kondo correlations may give rise to a more complicated lineshape~\cite{Paaske06}.

The inelastic cotunneling step corresponding to excitations from the singlet ground state to the triplet excited states emerges from the left diamond edges in Fig.~\ref{fig:diamonds}(a) (seen as an abrupt color change from white to blue). Because of the gate dependence of the singlet-triplet splitting (due to the gate dependence of $J_{12}$), the inelastic cotunneling step acquires a finite slope and crosses zero bias close to the center of the Coulomb diamond. At this point the double dot ground state switches from singlet to triplet. To the right of this crossing a cotunneling step again emerges, corresponding to the opposite excitation (from the triplet, which is now the ground state, to the singlet). This step ends when it intersects the right diamond edge. Also emerging from the zero-bias crossing is a weak conductance peak (for $V>0$) or conductance dip (for $V<0$), marked with red arrows, which signals the finite-bias position of the singlet-triplet transition (the slope is a result of the combined $V$ and $V_g$ dependencies of $J_{12}$). This "singlet-triplet line", which was experimentally observed in Refs.~\onlinecite{Osorio10, Roch08a}, is thus a signature of the ground state transition as a function of the applied voltages and appears since the elastic cotunneling conductance of the triplet is somewhat larger than that of the singlet (the lineshape, and the conditions under which it is visible, are discussed in detail in Sect.~\ref{sec:kondo}).
%Note that the zero-bias conductance is larger to the right of the transition point than to the left, i.e., the triplet ground state
%conducts better than the singlet.
At the points where the singlet-triplet line intersects the edge of the Coulomb diamond, there is a discontinuous change in the slope of the diamond edges. The reason is that when crossing this line, the molecular ground state is changed into one with a different voltage dependence, which directly determines the diamond slopes.

Figure~\ref{fig:diamonds}(b) shows the same conductance map, but under the influence of a magnetic field. Those SET resonances which correspond to excitations with a finite total spin are now split by the Zeeman effect. In addition, the Zeeman splitting of the triplet state results in a splitting of the corresponding inelastic cotunneling steps. When the ground state is a singlet, excitations to the three triplet states lead to three distinct gate-dependent cotunneling steps. To the right of the ground state crossing point, where the ground state is the $T_{-1}$ triplet, excitations to the singlet give rise to a gate-dependent cotunneling step and excitations to the $T_{0}$ triplet result in a gate-independent step (excitations from the $T_{-1}$ to the $T_{+1}$ triplet are suppressed by spin-selection rules for cotunneling, $\Delta S_{z} = 0, \pm 1$). In an experiment, the Zeeman splitting of the inelastic cotunneling steps provides an important check of the spin assignment of the excitations.

In both Figs.~\ref{fig:diamonds}(a) and (b), weak conductance peaks with the same gate dependence as the Coulomb diamonds are seen inside the Coulomb blockade region, close to the diamond edges. These peaks result from SET transitions starting from excited states~\cite{Schleser05}, which can be populated in a nonequilibrium situation due to inelastic cotunneling processes.
%Such cotunneling-assisted SET (CAST) has previously been observed in experiments on quantum dots~\cite{Schleser05} and suspended carbon nanotubes~\cite{Huettel09}.

\section{Lineshape of the singlet-triplet transition}\label{sec:kondo}

\begin{figure}
\begin{center}
\subfigure[]
{
\centering
\label{fig:FigExpCuts}
\includegraphics[width=1.0\columnwidth]{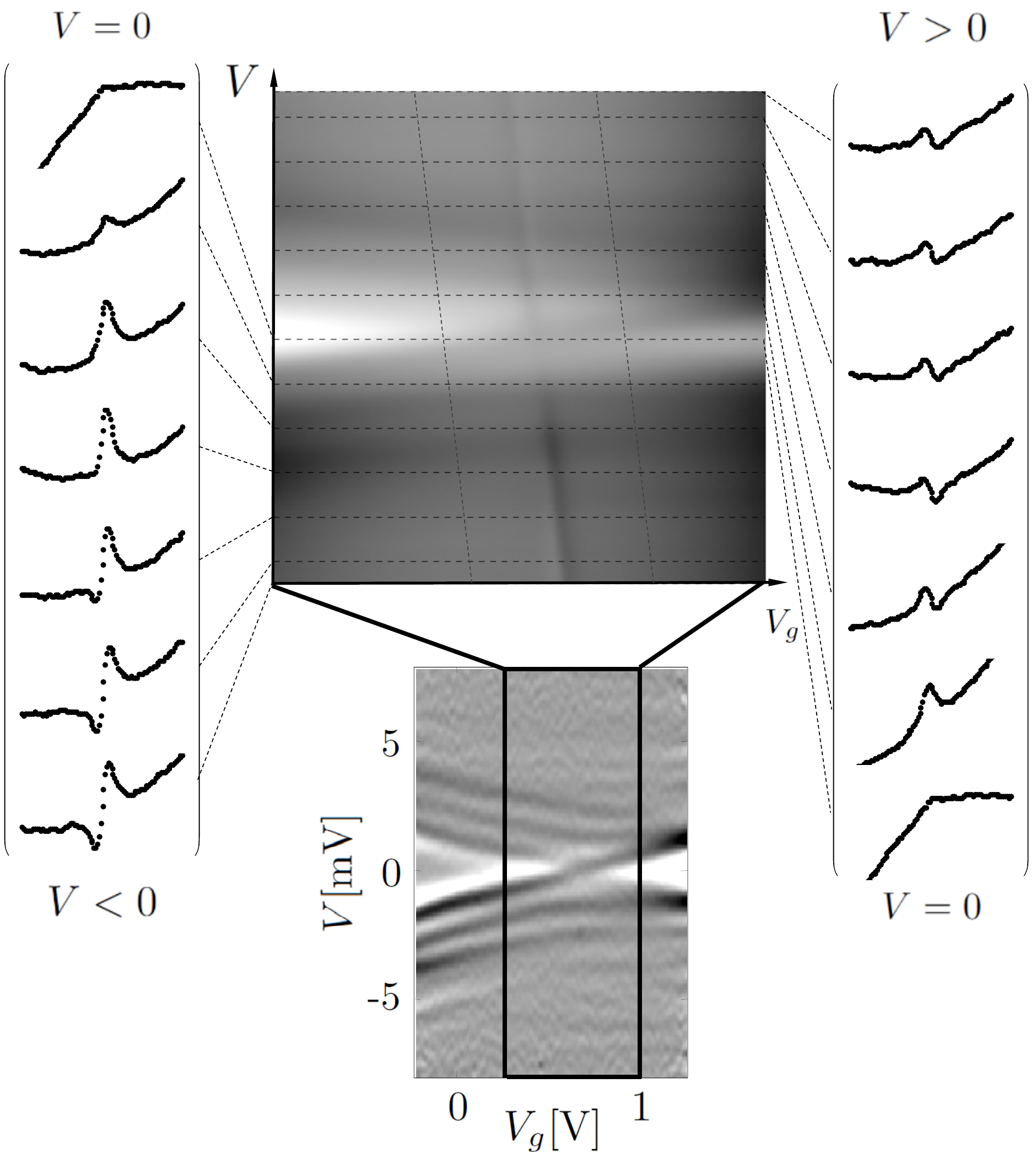}
}
\subfigure[]
{
\centering
\label{fig:DensPlot}
\includegraphics[width=1.0\columnwidth]{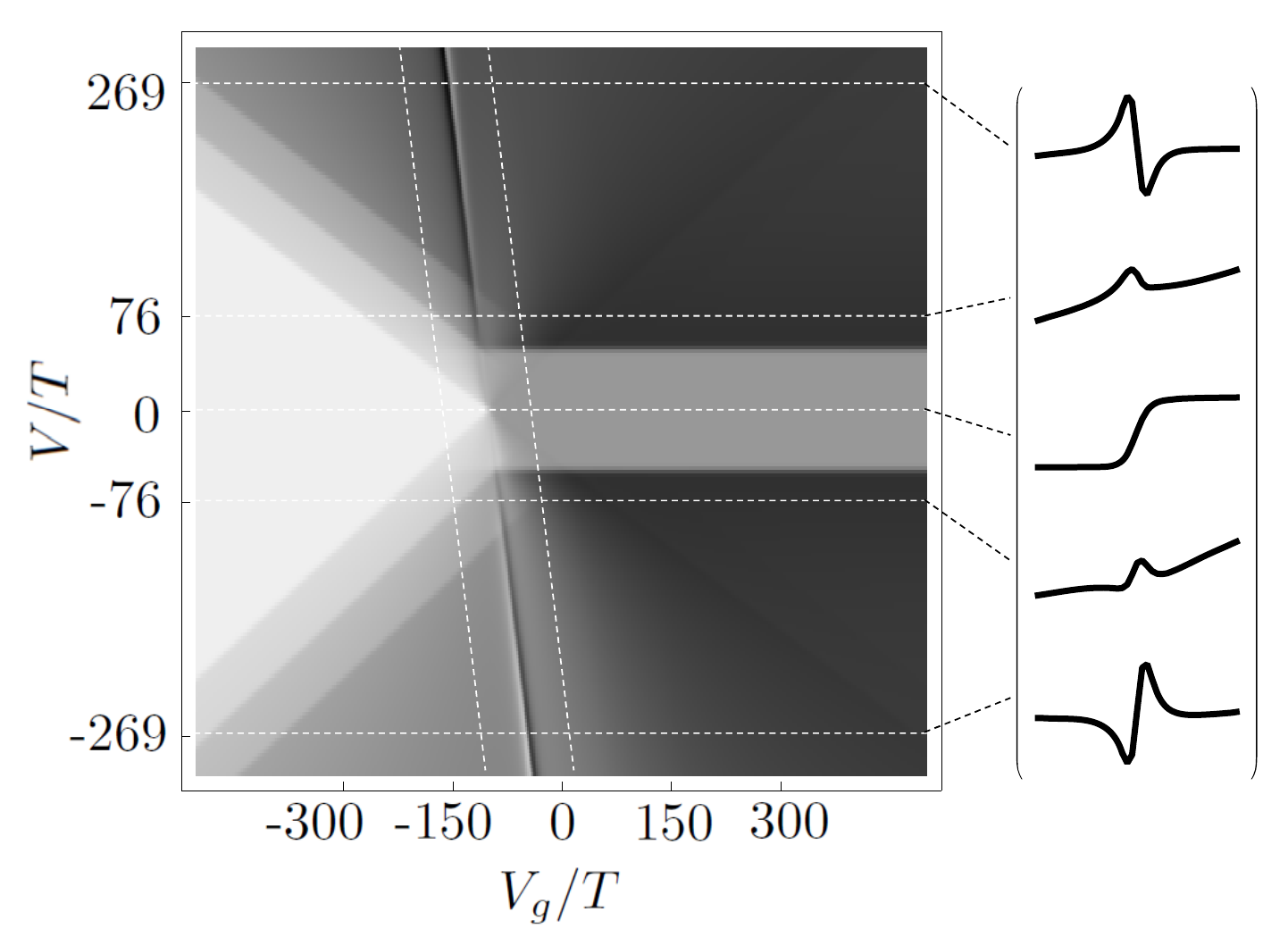}
}
\caption{(a) The lower density plot shows $d^{3}I/dV^{3}$ vs. bias voltage, $V$, and gate voltage, $V_{g}$, for the Mn complex measured in Ref.~\onlinecite{Osorio10}. An external magnetic field of 10 Tesla clearly reveals a singlet groundstate with an excited triplet towards the left, and a triplet groundstate with an excited singlet towards the right. The zoom-in shows the same data now plotted as $dI/dV$. A steep faint line is seen to be pinned to the groundstate crossing. Cuts at various bias voltages, marked by dashed lines, reveal a distinct lineshape as a function of $V_g$ developing at large applied bias: {\it dip-peak} at $V<0$ and {\it peak-dip} at $V>0$. The two nearly vertical dashed lines indicate the range in $V_g$ used for each of the cuts. The vertical range is the same in all the cuts, and has been centered at the mean value within each cut. (b) Density plot of differential conductance as calculated below in Sect.~\ref{sec:STmix}, with cuts at bias voltages $V/T=\pm269,\pm76,0$ (parameters are chosen as in Fig.~\ref{fig:symcutsV}). The peak-dip (dip-peak) structure from panel (a) is clearly reproduced. All density plots are on a grey scale where darker indicates higher values.}
\end{center}
\end{figure}

In Fig.~\ref{fig:diamonds}, the singlet-triplet line appeared as a peak or a dip in the conductance, for respectively positive or negative bias voltage. This was also the result of the calculations presented in the supplementary information to Ref.~\onlinecite{Osorio10}, but as pointed out already there, the experimental data from Ref.~\onlinecite{Osorio10}, reproduced here in Fig.~\ref{fig:FigExpCuts}, do not adhere to this behavior. Instead, the experiment shows a conductance peak, for both positive and negative bias voltage, which then evolves into a peak-dip, for $V>0$, or dip-peak, for $V<0$, roughly when the magnitude of the bias voltage becomes larger than all the relevant inelastic cotunneling thresholds. In this Sect., we will show that these Fano-like lineshapes along the singlet-triplet boundary can be reproduced by introducing a small mixing of the competing singlet and triplet states. In contrast to the scenario in Fig.~\ref{fig:diamonds} and the one discussed in the supplementary information to Ref.~\onlinecite{Osorio10}, this effect relies entirely on nonequilibrium pumping of excited states and therefore shows up only in devices with not too different couplings to source and drain electrodes.

\subsection{Effective exchange cotunneling model}

We now focus our attention to the region deep inside the Coulomb diamond with $N=2$, where real charge fluctuations (SET) are suppressed, and second order processes (cotunneling), involving virtual fluctuations to the nearby $N=3$ and $N=1$ charge states, dominate. If $U_i \gg U_{12}$, there is always one electron in each dot when $N = 2$. In this limit, the system described by the Hamiltonian~(\ref{eq:HD}) is effectively a two-spin system, and it is natural to work in the singlet-triplet basis $\{| T_{+1}\rangle,| T_{0}\rangle,| T_{-1}\rangle,| S\rangle\}$. In this basis, the {\it double-dot spin} Hamiltonian reads:
\begin{equation}\label{eq:HD2}
 H_{DDS}=\sum_{\eta}|\eta\rangle\varepsilon_{\eta}\langle\eta|,
%\left(
%\begin{array}{cccc}
% B+\frac{1}{4}J_{12} & 0 & 0 & 0 \\
% 0 & \frac{1}{4}J_{12} & 0 & 0 \\
% 0 & 0 & -B+\frac{1}{4}J_{12} & 0 \\
% 0 & 0 & 0 & -\frac{3}{4}J_{12}
%\end{array}
%\right),
\end{equation}
with eigenenergies
\begin{align}
\varepsilon_{S}=&-\frac{3}{4}J_{12},\\
\varepsilon_{T_{-1}}=&-B+\frac{1}{4}J_{12},\\
\varepsilon_{T_{0}}=&\frac{1}{4}J_{12},\\
\varepsilon_{T_{1}}=&B+\frac{1}{4}J_{12},
\end{align}
where we have omitted the constant term $U_{12} + \epsilon_1 + \epsilon_2$. As discussed above, $J_{12}$ is assumed to depend linearly on $V$ and $V_g$, and taking the singlet-triplet crossing point to be the zero of
our gate voltage axis, we thus assume an exchange coupling of the form:
\begin{align}\label{eq:Jfct}
	J_{12}=-\alpha V_g - \beta V,
\end{align}
in terms of two dimensionless numbers, $\alpha$ and $\beta$.

\begin{figure}
\includegraphics[width=0.9\columnwidth]{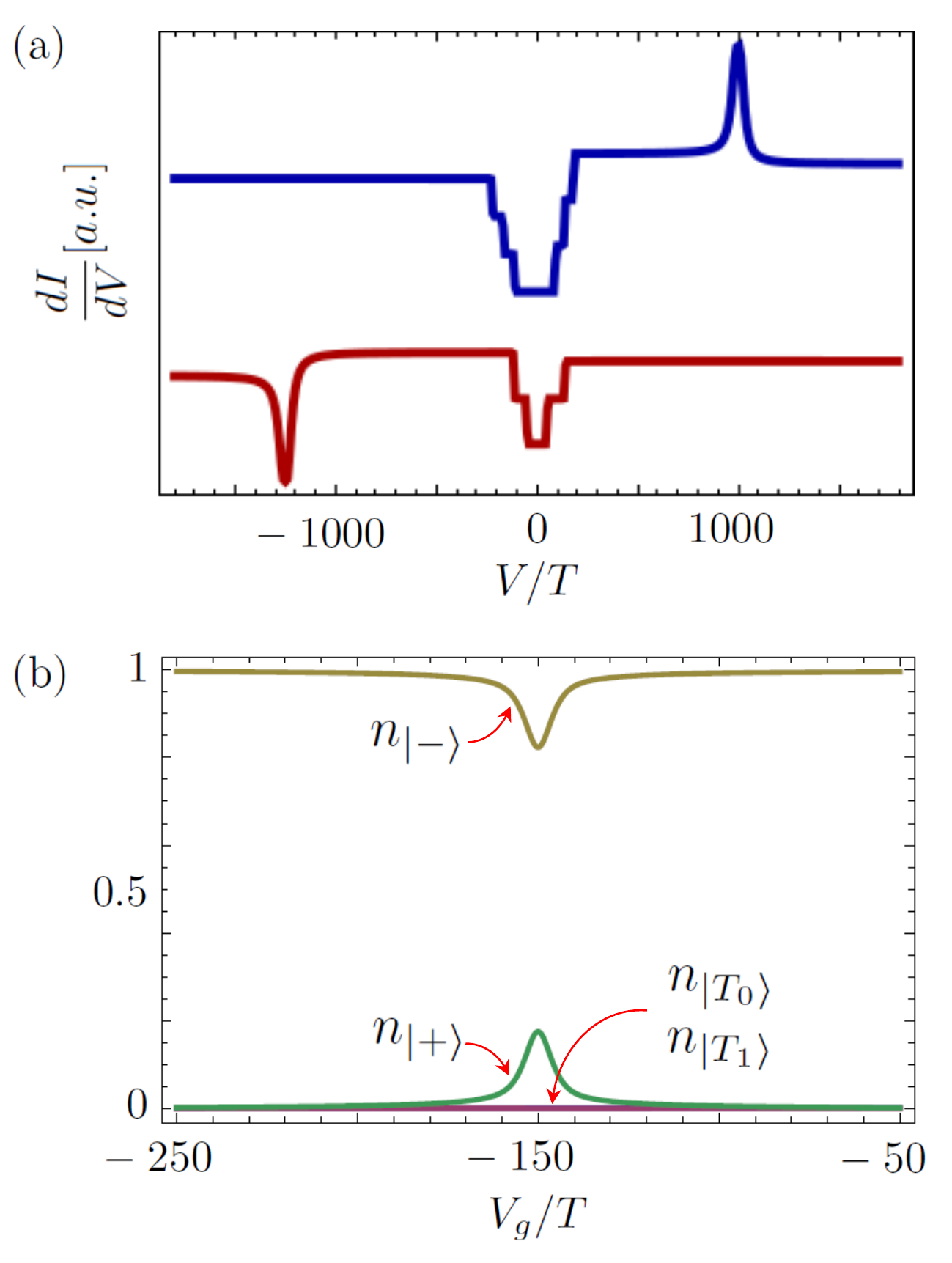}
\caption{\label{fig:asymcuts}(a) Differential conductance $dI/dV$ vs. $V/T$ at $V_{g}/T=-300$ (blue upper line) and at $V_g/T=200$ (red lower line). The curves are offset for clarity. The tunneling amplitudes are very asymmetric, $t_{1R} = 30, t_{1L} = 0.7, t_{2R} = 0.8, t_{2L} = 0.5$ (arbitrary units). We have furthermore chosen
$B/T = 50$, $\alpha = 0.5$, and $\beta = 0.1$.
(b) Occupation probabilities of the different spin states at fixed bias $V/T=250$ with parameters as in (a). The state $|+\rangle$ refers to the lowest lying state, i.e. $|S\rangle$ on the l.h.s. and $|T_{-1}\rangle$ on the r.h.s.. $|-\rangle$ refers to the next lowest, i.e. $|T_{-1}\rangle$ on the l.h.s. and $|S\rangle$ on the r.h.s.}
\end{figure}

\begin{figure}
\includegraphics[width=0.9\columnwidth]{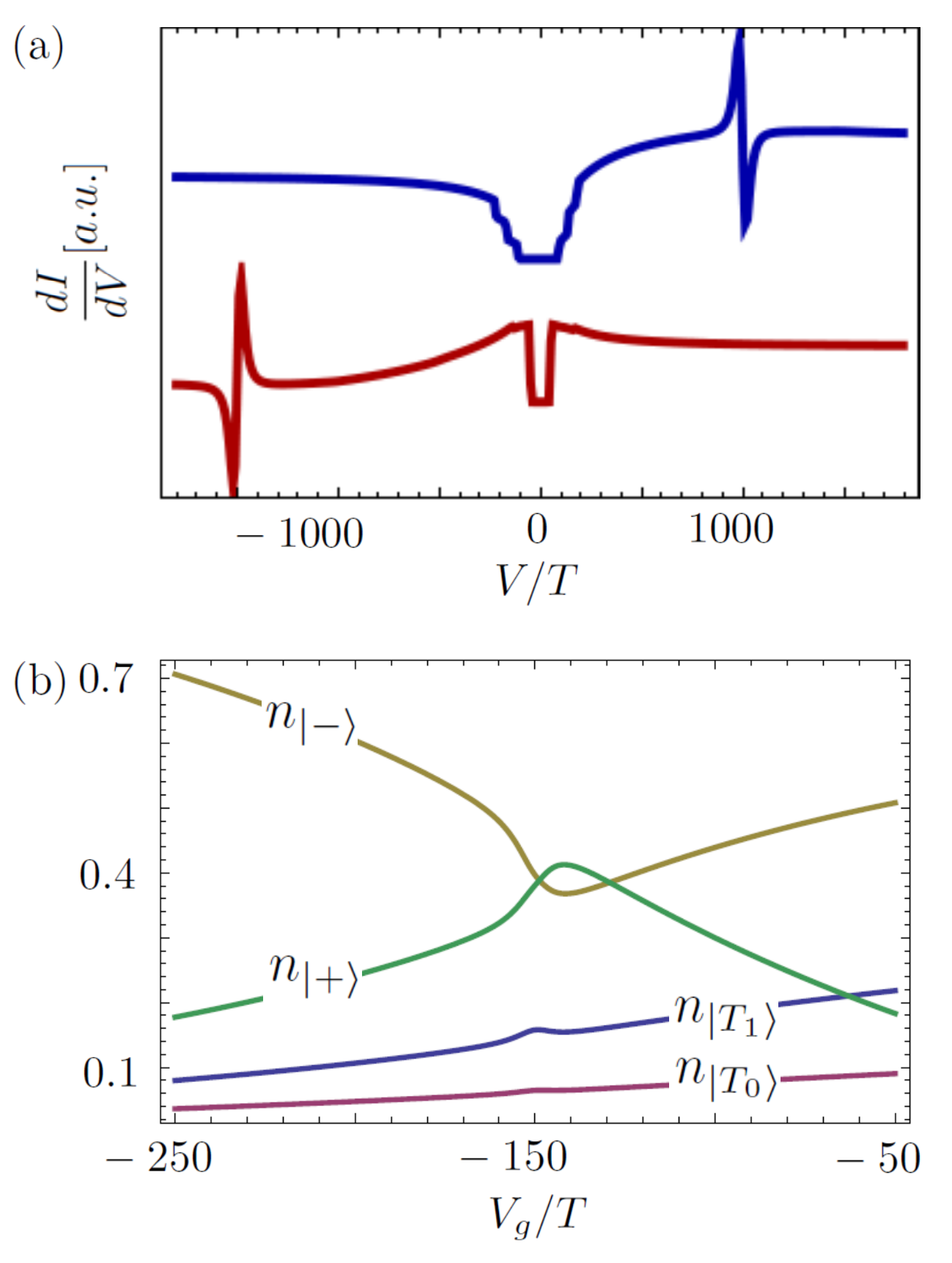}
\caption{\label{fig:symcutsV}(a) Differential conductance $dI/dV$ vs. $V/T$ at $V_{g}/T=-300$ (blue upper line) and at $V_g/T = 200$ (red lower line). The curves are offset for clarity. The tunnel amplitudes are now chosen more symmetric than in Fig.~\ref{fig:asymcuts}, with $t_{1R} = 0.3$,
%, t_{1L}/T = 0.7, t_{2R}/T = 0.8, t_{2L}/T = 0.5$
and a finite spin-orbit coupling, $D/T=2\sqrt{2}$ has been included. All other parameters are the same as in Fig.~\ref{fig:asymcuts}.
(b) Corresponding occupation probabilities of the different spin states at fixed bias $V/T=250$. Notice the occupation inversion between the two states $|-\rangle$ and $|+\rangle$ at their would-be degeneracy point at $V_{g}/T=-150$.}
\end{figure}

The kinetic energy of the electrode-electrons is still described by $H_R$ and, applying a Shrieffer-Wolff transformation~\cite{schriefferWolff66} to effectively eliminate the charge fluctuations, we arrive at the following exchange cotunneling Hamiltonian:
\begin{align}\label{eq:Hcot}
H_{cot}&=\!\!\sum_{\substack{r,r'=R,L\\i=1,2\\\sigma\sigma'=
\uparrow,\downarrow\\k,k'}}
\!\left(
J_{i,rr'}\bm{S}_{i}\cdot\bm{\tau}_{\sigma\sigma'}
+\frac{W_{i,rr'}}{2}\delta_{\sigma \sigma'}\right)
c^{\dag}_{r k\sigma}c_{r' k'\sigma'}.
\end{align}
where $\bm{\tau}$ is the vector of Pauli matrices. Written in the singlet-triplet basis, the local spin-operators,
\begin{align}
\bm{S}_{i} =\left((S_{i}^+ + S_{i}^-)/2, -i (S_{i}^+ - S_{i}^-)/2, S_{i}^z\right),
\end{align}
take the following form:
\begin{align}
S^{+}_{i}&=\frac{1}{\sqrt{2}}\left(
 \mid T_{+1}\rangle\langle T_{0}\mid
+\mid T_{0}\rangle\langle T_{-1}\mid
\right.\label{eq:Sop1}\\&\left.\hspace*{14mm}
\pm\mid T_{+1}\rangle\langle S\mid
\mp\mid S\rangle\langle T_{-1}\mid
\right),\nonumber\\
S^{-}_{i}&=\frac{1}{\sqrt{2}}\left(
\mid T_{0}\rangle\langle T_{+1}\mid
+\mid T_{-1}\rangle\langle T_{0}\mid
\right.\\&\left.\hspace*{14mm}
\pm\mid S\rangle\langle T_{+1}\mid
\mp\mid T_{-1}\rangle\langle S\mid
\right),\nonumber\\
S^{z}_{i}&=\frac{1}{2}\left(
\mid T_{+1}\rangle\langle T_{+1}\mid
-\mid T_{-1}\rangle\langle T_{-1}\mid
\right.\label{eq:Sop3}\\&\left.\hspace*{14mm}
\mp\mid S\rangle\langle T_{0}\mid
\mp\mid T_{0}\rangle\langle S\mid
\right).\nonumber
\end{align}
The spin exchange, and the potential-scattering amplitudes are related to the tunneling amplitudes appearing in~\eqref{eq:HT} as:
\begin{align}
 J_{i,rr'} &=t^{*}_{ir}t_{ir'}\left(\frac{1}{E_2-E_3}+\frac{1}{E_2-E_1}\right),\\
 W_{i,rr'} &=t^{*}_{ir}t_{ir'}\left(\frac{1}{E_2-E_3}-\frac{1}{E_2-E_1}\right),
\end{align}
where the energy denominators keep track of the energy cost for the virtual charge fluctuations to respectively the $N=3$ and the $N=1$ states. Assuming that we are deep inside the Coulomb diamond, we have omitted the slight differences in these denominators for the different spin-states. The potential scattering term vanishes close to the center of the diamond, where $E_3-E_2=E_1-E_2$. Since this gate voltage dependence has little influence on the line shape of the singlet-triplet crossing studied below, we shall assume that $|J_{i,rr'}| = |W_{i,rr'}|$. This has been done in the results shown in Figs.~\ref{fig:asymcuts} and~\ref{fig:symcutsV}.

The cotunneling current can now be found as~\cite{Bruus04book}
\begin{align}\label{eq:curr}
I=e\sum_{\eta'\eta}(\Gamma^{RL}_{\eta'\eta}
-\Gamma^{LR}_{\eta'\eta})P(\eta).
\end{align}
The non-equilibrium occupation numbers for the four singlet-triplet basis states, $P(\eta)$, are found by solving the rate equations
\begin{align}
0=\sum_{\eta'}\left(
\Gamma_{\eta'\eta}P(\eta)-\Gamma_{\eta\eta'}P(\eta')\right),
\end{align}
with the constraint that
\begin{align}
1=\sum_{\eta}P(\eta),
\end{align}
and with rates
\begin{align}
\Gamma_{\eta'\eta}=\sum_{r'r}
\Gamma^{r'r}_{\eta'\eta},
\end{align}
determined from Fermi's golden rule as
\begin{align}
\Gamma^{r'r}_{\eta'\eta}=
\gamma^{r'r}_{\eta'\eta}
\Delta\varepsilon^{r'r}_{\eta'\eta}
n_{B}(\Delta\varepsilon^{r'r}_{\eta'\eta}).
\end{align}
Here $n_{B}$ denotes the Bose function, and we have introduced the energy differences
\begin{align}
\Delta\varepsilon^{r'r}_{\eta'\eta}=
\varepsilon_{\eta'}-\varepsilon_{\eta}+\mu_{r}-\mu_{r'},
\end{align}
and the tunneling probabilities
\begin{align}
\gamma^{r'r}_{\eta'\eta}=&
\pi\rho^{2}\sum_{i=1,2}\left(\sum_{j=x,y,z}
|J_{i,r'r}|^{2}|T_{i,\eta'\eta}^{j}|^{2}
%\right.\nonumber\\&\hspace*{22mm}\left.
+|W_{i,r'r}|^{2}\delta_{\eta'\eta}\right),
\end{align}
where the matrix elements $T_{i,\eta'\eta}^{j}=\langle\eta'|S^{j}_{i}|\eta\rangle$ can be read off from Eqs. (\ref{eq:Sop1})--(\ref{eq:Sop3}).

If the device is coupled much stronger to one electrode than the other, the nonequilibrium occupations of excited states are strongly suppressed, and the finite-bias conductance becomes a simple succession of steps at bias voltages corresponding to energies of the excited states. Already from the elastic cotunneling amplitude, $\varGamma^{RL}_{\eta\eta}$, one can see that the conductance is larger for a triplet than for a singlet ground state
\begin{align}
\varGamma^{RL}_{SS}=&4\pi\rho^{2} V n_{B}(V)|W_{RL}|^{2},\\
\varGamma^{RL}_{T_{-1}T_{-1}}=&4\pi\rho^{2} V n_{B}(V)\left(|J_{RL}|^{2}+|W_{RL}|^{2}\right),
\end{align}
which amounts to a factor of two in difference between conductance on the singlet, or the triplet side if $|J_{i,rr'}|=|W_{i,rr'}|$. If the singlet-triplet boundary is completely vertical in the $(V_g,V)$-plane ($J_{12}$ depends only on $V_g$), this difference in conductance only shows up as a change in contrast in the conductance map as indicated in Fig.~\ref{fig:LevCross}. If, on the other hand, the singlet-triplet crossing has a slight bias-dependence, this already explains why Fig.~\ref{fig:diamonds} shows a peak for positive, and a dip for negative bias voltage. This is seen clearly in Fig.~\ref{fig:asymcuts}(a), which shows $dI/dV$ as a function of $V$ for two different values of $V_g$, calculated from formula \eqref{eq:curr}. The singlet-triplet crossing gives rise to a peak at positive bias (blue line) and a dip at negative bias (red line). Figure~\ref{fig:asymcuts}(b), shows the corresponding nonequilibrium occupation numbers, and since we have chosen a much stronger coupling to the right electrode, nonequilibrium effects are weak and one observes only slight changes in a small region about the crossing.

\subsection{Including singlet-triplet mixing}\label{sec:STmix}

The cotunneling spectroscopy of the Mn complex in Ref.~\onlinecite{Osorio10} reveals a transition between the lowest lying and the highest lying triplet states (cf. r.h.s. of lower panel of Fig.~\ref{fig:FigExpCuts}). This implies a violation of the normal spin-selection rules for cotunneling, from which we infer that the relevant spin-states are not really true eigenstates of $S^{z}$, but instead a mixture of these inflicted by spin-orbit coupling in the molecule. We now add this information to our model by including a simple Dzyaloshinskii-Moriya term~\cite{Dzyaloshinsky58a,Moriya60a,Herzog10}, $H_{DM}=({\bf S}_{1}\times{\bf S}_{2})\cdot{\bf D}$, which reads
\begin{align}
H_{DM}=\frac{D}{2\sqrt{2}}\left[(|T_{+1}\rangle +|T_{-1}\rangle)\langle S|+|S\rangle(\langle T_{+1}| + \langle T_{-1}|)\right],
\end{align}
for ${\bf D}\perp{\bf B}$. This has the important effect of mixing the competing singlet and triplet states and will in general give rise to an avoided crossing as illustrated in Fig.~\ref{fig:LevCross}, with $\lambda\sim D$. If, however, the coupling strength $D$ is smaller than the bias width of the cotunneling peaks, this avoided crossing need not be observed. Interestingly, even such a small spin-orbit mixing may still have an observable effect on the line shape of the rather steep finite-bias singlet-triplet boundary. For simplicity we only include the part of $H_{DM}$ which mixes the singlet and lowest triplet ($T_{-1}$) states, which is the only relevant term close to the singlet-triplet line.

Including $H_{DM}$, we may determine the new eigenstates at a given bias voltage, $V$, which still enters via $J_{12}(V_g,V)$. Using the same procedure as outlined above, the cotunneling current is readily recalculated. The result obtained for $D/T=2\sqrt{2}$ is shown in Fig.~\ref{fig:symcutsV}(a), revealing a pronounced peak-dip, and dip-peak structure at respectively positive and negative bias, in qualitative correspondence with the experimental observation. Whereas the peak (dip) in Fig.~\ref{fig:asymcuts}, with no spin-orbit coupling included, could only be obtained with a substantial difference in the couplings to respectively source and drain electrodes, the peak-dip (dip-peak) structure in Fig.~\ref{fig:symcutsV} only arises when the couplings to source and drain are not too different. The observed peak-dip structure is thus a nonequilibrium effect arising from the bias dependent occupation numbers of the various spin-states, shown in Fig.~\ref{fig:symcutsV}(b) to exhibit a population inversion of the two competing ground states near the crossing, around $V_g/T\sim -150$ in this plot. Apart from the requirement of nearly equal tunnel couplings, this population inversion and the concomitant peak-dip structure does not rely on any fine-tuning of parameters. Fig.~\ref{fig:DensPlot} shows this result in a full density plot of the nonlinear conductance with a number of gate voltage traces taken at different bias voltages, all plotted in the same style as the experimental data in Fig.~\ref{fig:FigExpCuts}. The characteristic Fano-like lineshape is seen both when the conductance is plotted as a function of $V$ and of $V_g$.

Reading off the slope of the nearly vertical singlet-triplet boundary and the slope of the inelastic singlet-triplet cotunneling threshold in zero magnetic field, one can now readily determine the constants $\alpha$ and $\beta$ from formula \eqref{eq:Jfct}. Doing this, we find that $\alpha\approx 0.002$ and $\beta\approx 0.020$. In Fig.~\ref{fig:FigExpCuts}, the width in gate voltage of the peak-dip structure is roughly $0.05$ V, and using $\alpha$ this now translates into a corresponding change in $J_{12}$ of the order of 0.1 meV, which must be an indirect measure of the effective singlet-triplet mixing, $D$, in this Mn-complex. 
%This is roughly half of the experimental temperature, $T=1.7$ K, and thus tentatively consistent with the fact that no avoided crossing is observed at zero bias voltage. Whereas the atomic spin-orbit coupling of Mn is of the order of 20-30 meV~\cite{Bendix93}, it is hard to say what the effective singlet-triplet mixing is for the states probed in the transport measurement of the Mn-complex, especially since one of the two spins is believed to reside on the terpyridine moity of the molecule (cf. Ref.~\onlinecite{Osorio10}).

\section{Conclusions}

Inspired by the experimental results for the conductance map of a Mn complex reported in Ref.~\onlinecite{Osorio10}, we have studied the signatures in the nonlinear conductance near a finite-bias singlet-triplet transition. For asymmetrically coupled devices the nonequilibrium effects are weak. A mere difference in elastic cotunneling amplitudes for respectively singlet and triplet states can give rise to current plateaus of different heights and hence a conductance peak along the transition line in the $(V,V_{g})$-plane. For more symmetrically coupled devices, this effect will be washed out at sufficiently large bias voltages. In this case, however, a new possibility arises, if the two competing states are mixed by e.g. spin-orbit coupling. In this case, the nonequilibrium occupations exhibit a momentary population inversion near the transition line, which in turn gives rise to a peak-dip anomaly in the conductance. This provides an explanation for the observations in Ref.~\onlinecite{Osorio10}, and it provides an indirect method of gauging the size of the singlet-triplet mixing.

This type of cotunneling signature should be rather general, and the considerations about the singlet-triplet transition studied here could readily be exported to entirely different Coulomb blockaded systems with level crossings induced by gate-voltage. If the competing states are sensitive to gate voltage, it is only natural to expect also a slight sensitivity to bias voltage and thus a non-vertical transition line as depicted in Fig.~\ref{fig:LevCross}. As was the case for the Mn complex discussed here, the steepness of this line can help bring out otherwise hidden spectroscopic details.

\section{acknowledgements}
The research leading to these results has received funding from the European Union Seventh Framework Programme
(FP7/2007-2013) under agreement no 270369 (“ELFOS”).

\bibliographystyle{apsrev}
%\bibliography{cite}
%Merlin.mbs v4.21 2009-07-09.
%
\end{document}